\documentclass[preprint,prl,superscriptaddress,amsmath,amssymb]{revtex4}

\usepackage{bm}
\usepackage{color}
\usepackage[dvips]{graphicx}

\makeatletter

\def\lsim{\compoundrel<\over\sim}
\def\compoundrel#1\over#2{\mathpalette\compoundreL{{#1}\over{#2}}}
\def\compoundreL#1#2{\compoundREL#1#2}
\def\compoundREL#1#2\over#3{\mathrel
    {\vcenter{\hbox{$\m@th\buildrel{#1#2}\over{#1#3}$}}}}
\makeatother

\begin{document}

\title{Three-dimensional bulk band dispersion in polar BiTeI with giant Rashba-type spin splitting}

\author{M. \surname{Sakano} }
\affiliation{Department of Applied Physics, The University of Tokyo, Tokyo 113-8656, Japan}

\author{J. \surname{Miyawaki} }
\affiliation{RIKEN SPring-8 Center, Sayo-gun, Hyogo, 679-5148, Japan}

\author{A. \surname{Chainani} }
\affiliation{RIKEN SPring-8 Center, Sayo-gun, Hyogo, 679-5148, Japan}

\author{Y. \surname{Takata} }
\affiliation{RIKEN SPring-8 Center, Sayo-gun, Hyogo, 679-5148, Japan}

\author{T. \surname{Sonobe} }
\affiliation{Department of Applied Physics, The University of Tokyo, Tokyo 113-8656, Japan}

\author{T. \surname{Shimojima} }
\affiliation{Department of Applied Physics, The University of Tokyo, Tokyo 113-8656, Japan}

\author{M. \surname{Oura} }
\affiliation{RIKEN SPring-8 Center, Sayo-gun, Hyogo, 679-5148, Japan}

\author{S. \surname{Shin} }
\affiliation{RIKEN SPring-8 Center, Sayo-gun, Hyogo, 679-5148, Japan}
\affiliation{Institute of Solid State Physics, The University of Tokyo, Kashiwa, Chiba, 227-8581, Japan}
\affiliation{CREST,JST, Tokyo 102-0075, Japan}

\author{M. S. \surname{Bahramy} }
\affiliation{Correlated Electron Research Group (CERG), RIKEN Advanced Science Institute, Wako, Saitama 351-0198, Japan}

\author{R. \surname{Arita} }
\affiliation{Department of Applied Physics, The University of Tokyo, Tokyo 113-8656, Japan}
\affiliation{Correlated Electron Research Group (CERG), RIKEN Advanced Science Institute, Wako, Saitama 351-0198, Japan}
\affiliation{JST, PRESTO, Kawaguchi, Saitama 332-0012, Japan}

\author{N. \surname{Nagaosa} }
\affiliation{Department of Applied Physics, The University of Tokyo, Tokyo 113-8656, Japan}
\affiliation{Correlated Electron Research Group (CERG), RIKEN Advanced Science Institute, Wako, Saitama 351-0198, Japan}

\author{H. \surname{Murakawa} }
\affiliation{Correlated Electron Research Group (CERG), RIKEN Advanced Science Institute, Wako, Saitama 351-0198, Japan}

\author{Y. \surname{Kaneko} }
\affiliation{Multiferroics Project, ERATO, JST, Tokyo 113-8656, Japan}

\author{Y. \surname{Tokura} }
\affiliation{Department of Applied Physics, The University of Tokyo, Tokyo 113-8656, Japan}
\affiliation{Correlated Electron Research Group (CERG), RIKEN Advanced Science Institute, Wako, Saitama 351-0198, Japan}
\affiliation{Multiferroics Project, ERATO, JST, Tokyo 113-8656, Japan}

\author{K. \surname{Ishizaka} }
\affiliation{Department of Applied Physics, The University of Tokyo, Tokyo 113-8656, Japan}
\affiliation{JST, PRESTO,  Kawaguchi, Saitama 332-0012, Japan}

\begin{abstract}
In layered polar semiconductor  BiTeI, giant Rashba-type spin-split band dispersions show up due to the crystal structure asymmetry and the strong spin-orbit interaction.
Here we investigate the 3-dimensional (3D) bulk band structures of BiTeI  using the bulk-sensitive $h\nu$-dependent soft x-ray angle resolved photoemission spectroscopy (SX-ARPES).
The obtained  band structure is shown to be well reproducible by the first-principles calculations, with huge spin splittings of ${\sim}300$ meV at the conduction-band-minimum and valence-band-maximum located in the $k_z=\pi/c$ plane. 
It provides the first direct experimental evidence of the 3D Rashba-type spin splitting in a bulk compound.
\end{abstract}

\pacs{71.45.Lr, 78.47.J-, 79.60.-i}
\maketitle


Symmetry of the crystal lattice plays an essential role to determine the electronic structures of solids. Among them, the spatial inversion $I$ is an important symmetry since it connects the momenta $k$ and $-k$, which are related also by the time-reversal symmetry $T$. In the latter case, the spin is also reversed, and hence the interplay of $I$- and $T$ symmetries in the presence of the spin-orbit interaction (SOI) is an important issue. 
Investigation of SOI in noncentrosymmetric materials originates in Dresselhaus and Rashba effects \cite{Dresselhaus55,Rashba60}, which evaluate the spin-split band structures in zincblende and wurtzite semiconductors. 
More recently, the Bychkov-Rashba model \cite{Bychkov84} was proposed as a suitable explanation of the properties of the 2-dimensional electron gas with perpendicular electric field $\vec E_z$ applied. 
In this model, the electron with momentum $\vec{k}$ and spin $\vec{\sigma}$ feels an effective magnetic field perpendicular to $\vec k$ and $\vec E_z$, as represented by the Hamiltonian $H_{\rm{BR}}=\lambda \, \vec{\sigma} \cdot (\vec E_z \times \vec{k})$, with coupling constant $\lambda$. 
Such SOI effects in inversion asymmetric systems introduce various spin-related electric properties, such as spin Hall effect \cite{Hirsch99} and spin-galvanic current \cite{Manchon08,Chernysnov09,Miron10}, and consequently, occupy an important position in the spintronics field. 
Unconventional quantum electronic phase is also expected to appear in noncentrosymmetric superconductors, induced by the mixing of spin-singlet and spin-triplet states \cite{Edel'stein89,Gor'kov01}. 
From the viewpoint of elucidating the SOI-induced novel phenomena and spintronic functions, the development of compounds with high-energy-scale spin-splitting is highly desirable. 

Pursuit for giant spin-splitting has been mainly focused on the Shockley state of single crystals composed of heavy elements until now \cite{LaShell96,Hoesch04}, which is characterized by the clear structural asymmetry at the topmost surface cutoff. 
Indeed, the huge Rashba-type splitting was observed in Bi/Ag(111) surface alloyed system \cite{Ast07}, where the $k$-linear spin splitting similar to the Bychkov-Rashba dispersion $E^{\pm}(\vec k ) = {\hbar} ^{2} \vec{k} ^{2} / 2m^{\ast} \pm \alpha_{\rm{R}} |\vec{k}|$ ($m^{\ast}$: effective mass of electron) had been demonstrated by angular-resolved photoemission spectroscopy (ARPES). 
Here, $\alpha_{\rm{R}}$ is the Rashba parameter indicating the strength of Rashba-type SOI, and $\alpha_{\rm{R}} = 3$ {eV\AA} 
is achieved in Bi/Ag(111) system. Most recently, on the other hand, the comparable size of spin splitting ($\alpha_{\rm{R}}=3.8$ eV\AA) has been found to be realized in a bulk polar semiconductor, BiTeI \cite{Ishizaka11}. 
Due to the peculiar trigonal polar structure ($P3m1$) arising from the alternative stacking of Bi- Te- and I- triangular networks [see Fig. 1(a)], it also shows a Rashba-type spin splitting similar to the ones observed in 2-dimensional surface Shockley states. 
Quantitative agreements between the (spin-resolved) ARPES and the first-principles calculations, together with the optical study on the bulk state, strongly suggest that the spin splitting is derived from bulk BiTeI \cite{Ishizaka11,Bahramy11,Lee11}. 
Nevertheless, the direct observation of 3-dimensional (3D) spin-splitting beyond Bychkov-Rashba model has not been investigated to date. 
Recent theoretical discovery of pressure-induced band inversion and topological transition in BiTeI \cite{Bahramy12} further increases the importance of directly confirming the bulk band structure in detail.
In this paper, we investigate the band structure of BiTeI throughout the whole 3D Brillouin zone, by utilizing the photon-energy ($h\nu$) dependent bulk-sensitive soft x-ray ARPES (SX-ARPES). 
The result gives the first direct experimental evidence of 3D Rashba-type spin-splitting, which leads to a deeper understanding of SOI in solids.  

Single crystals of  BiTeI were grown by Bridgman method \cite{Ishizaka11}. 
The electron-carrier density was estimated to be $4.5\times 10^{19}$ cm$^{-3}$ from the Hall coefficient, showing the $n$-type degenerate semiconducting behavior. 
$h\nu$-dependent SX-ARPES experiments were performed by using a Gammadata-Scienta SES2002 hemispherical electron analyzer at the undulator beam line BL17SU of SPring-8. 
The system adopts a grazing-incidence geometry ($<$5$^\circ$)  spectrometer that enables ARPES measurements to be highly efficient and allows an accurate determination of the $k_{x}-k_{y}$ momentum due to negligible modification of the $k_z$ momentum\cite{Ohashi07, Takata04}. 
The sample was cleaved {\it{in situ}} under an ultrahigh vacuum of $1{\times}10^{-8}$ Pa. 
All experiments were carried out at $50$ K. 
The Fermi level of the sample was referred to that of the polycrystalline Au electrically connected to the sample. 
In order to avoid the contamination of Auger lines, soft x-ray between $530$ eV and $590$ eV were used.
Both right and left circularly polarizations were used to avoid any polarization selection rule. 
The momentum region  of $h\nu$-dependent SX-ARPES measurement is shown in Fig. 1(b) together with the Brillouin zone of BiTeI. 
All measurements were performed in the red translucent plane, which nearly corresponds to  the $\Gamma$-M-L-A ($k_{x}-k_{z}$) plane. 
Figure 1(c) shows the momentum cuts of measurements we can access by using the photon energies of 530-590 eV at intervals of 5 eV. 
The red solid (blue broken) curves correspond  to the measurements in which the energy resolution was set to $100$ meV ($150$ meV). 

SX-ARPES images along $k_x$ direction are shown in Figs. 2(a)-(c) for $h\nu = 540$, 565 and 585 eV, which correspond to the $k_z$ values of 0, $0.56{\pi/c}$ and $\pi/c$, respectively [see Fig. 1(c)].
Peak positions of SX-ARPES intensity with error-bars estimated from momentum distribution curves (MDCs) or energy distribution curves (EDCs) are plotted as black-circle markers on SX-ARPES images, to show the dispersions of valence bands (VBs) below the band gap. 
At $k_{x} = 0$, six VBs are observed below $E_{\rm{F}}$, which mainly consist of Te and I $5p$ orbitals.
Some of them further show splitting in $k_{x}\neq 0$ region due to the SOI, as will be discussed later.
Observed band structures show clear $k_z$-dependence reflecting the 3D bulk electronic structure. 
Focusing on the highest valence band (HVB), its top lies at the binding energy ($E_{{B}}$) of $1.0$ eV at $k_z = 0$ while it moves upward and forms the VB maximum (VBM) of $E_{{B}} = 0.48$ eV at $k_{z} = {\pi/c}$. 
The conduction band (CB) above the band gap, on the other hand, seems to be located above $E_{\rm{F}}$ at $k_{z}=0$, but  gradually moves below the Fermi level and appears at around $k_{x} = 0$ for $k_{z} = 0.56{\pi/c}$ and $\pi/c$. 
The energy position of the conduction band minimum (CBM) is $E_{{B}}\sim 0.22$ eV at $k_z = \pi/c$.  
Thus in the bulk band structure of BiTeI, both the VBM and CBM are located within $k_{z} = \pi/c$ plane, forming the minimum gap of $\sim 0.26$ eV. 

To compare with the bulk band dispersions recorded by SX-ARPES, the results of the relativistic first-principles calculations \cite{comment1} for $k_{z} = 0$, 0.56$\pi/c$ and $\pi/c$ are shown in Figs. 2(e)-(g). 
The energy axis of the calculated data is rigidly shifted to match the energy of VBM at $E_{{B}} = 0.48$ eV obtained by SX-ARPES.
Calculated VBs consist of 12 bands because of spin splitting by SOI. 
All bands are spin degenerate along $\Gamma$-A line ($k_{x} = k_{y} = 0$) due to the time-reversal symmetry and the $P3m1$ symmetry of  the crystal.
Although some deeper VBs at $E_{{B}}$ = 2 $\sim$ 6 eV tend to have somewhat lower binding energies compared to the SX-ARPES data, the shapes of the calculated bands coincide well with those obtained experimentally.
The spin splittings for these deep VBs are also partly observed by SX-ARPES, {\it{e.g.}} the deepest VB showing the energy split of $\sim100$ meV around $k_{x}= \pm 0.1$ {\AA}$^{-1}$ in Fig. 2(c). 
The calculated CB bottom (HVB top), on the other hand, gets deeper (shallower) in binding energy with increasing $k_z$, and the minimum gap is formed at $k_{z} = {\pi/c}$. 
Simultaneously, the spin splittings of the CB and HVB become bigger, showing the huge band splitting of about $300$ meV realized at CBM and VBM in the $k_{z} = \pi/c$ plane.
These are found to show the $k_{\parallel}$-linear spin splitting in a small $k_{\parallel}$ region ($k_{\parallel}=\sqrt{k_x^2+k_y^2}$),  as similarly reported in wurtzite cases \cite{Rashba60,Casella60}. 
Also in SX-ARPES image at $k_{z} = \pi/c$ [Fig. 2(c)], the similar spin splitting of the HVB is clearly seen in the green rectangle, as will be discussed later quantitatively together with that of CB.

In Fig. 2(d), the band dispersions along $\Gamma$-A line obtained by SX-ARPES using $h\nu$ of 530 $\sim$ 590 eV are shown, to elucidate the $k_z$-dependence of the band structure. 
The calculation as shown in Fig. 2(h) accords well  with the band structure obtained by SX-ARPES.
The CB, HVB, and the deepest VB show moderately strong $k_{z}$-dependence ($0.4 \sim 0.8$ eV) as compared to other bands ($<0.2$ eV), since they are mainly characterized by Bi $6p_z$, Te $5p_z$, and I $5p_z$, respectively \cite{Bahramy11}.
Still they are apparently weaker than the in-plane dispersions.
This anisotropic electronic structure reflects the layered crystal structure of BiTeI, which is also confirmed by the anisotropic effective mass of $m_{z}/m_{x} \,{\sim}5$ by optical study \cite{Lee11}.
On the other hand, since the CB crosses $E_{\rm{F}}$ between $k_{z} = 0$ and $\pi/c$, it should show up as the $n$-type closed spin-split Fermi surface pockets centered at $k_z=\pi/c$, whose topology should be characterized by 3D-Rashba model \cite{Cappelluti07}.

To quantitatively compare the results of spin splitting obtained by SX-ARPES with that of the calculation, a close-up around the minimum gap, corresponding to the green-rectangle region in Fig. 2(c), is shown in Fig. 3(a).
The peak positions estimated by EDCs [shown in Fig. 3(b)]  and MDCs are represented as  circles and squares for HVB, while the guides for eyes are shown as green broken curves to trace CB.  
ARPES intensity of the CB is much weaker than that of the HVB in Figs. 3(a) and 3(b), because of the small photoabsorption cross-section of the Bi $6p$ orbitals. 
The relativistic first-principles calculations as shown in Fig. 3(c) indicates the Rashba-type spin splitting of CB and HVB due to the SOI under the $P3m1$ symmetry, which are very similar to the SX-ARPES results. 
To evaluate the strength of the Rashba-type spin splitting in analogy with the 2DEG Bychkov-Rashba model, we quantify the aforementioned Rashba parameter ${\alpha}_{\rm{R}}$.
Here we estimate ${\alpha}_{\rm{R}}$ by using the momentum offset at the CBM ($k_{0}^{\rm{CBM}}$), the binding energy of the CBM ($E_{\rm{CBM}}$) and the outer Fermi momentum ($k_{\rm{F}}^{\rm{out}}$). 
Most simply, by assuming the parabolic band dispersions, they are related by ${\alpha}_{\rm{R}}={{{\hbar}^2}k_{0}^{\rm{CBM}}/m_x}$ and $E_{\rm{CBM}}={\hbar}^{2}(k_{\rm{F}}^{\rm{out}}-k_{\rm{0}}^{\rm{CBM}})^2/{2m_x}$. 
Our SX-ARPES spectra at $k_z=\pi/c$ shows the ``W" like shaped CB with $|k_{0}^{\rm{CBM}}|=0.04 \pm 0.01$ {\AA}$^{-1}$, $|k_{\rm{F}}^{\rm{out}}| = 0.10 \pm 0.01$ {\AA}$^{-1}$ and $E_{\rm{CBM}} = 220 \pm 30$ meV. 
Thus we can evaluate the Rashba parameter as ${\alpha}_{\rm{R}} = 4.9 \pm 1.6$ eV{\AA}.
The value of ${\alpha}_{\rm{R}}$ lies within error bars of the value previously obtained from the CB in the surface accumulation layer of BiTeI, and gives one of the strongest Rashba-type spin splitting as reported.
Spin splittings can be also observed at the HVBs in Fig. 3(a). 
The offset momentum at the VBM is $|k_{0}^{\rm{VBM}}| = 0.05 \pm 0.01$ {\AA}$^{-1}$, whereas the spin splitting energy is estimated to be $\sim 220$ meV.
These are fairly similar in size with the splitting of CB as mentioned above.

This huge spin-splitting has been successfully explained by means of $k\cdot p$ perturbation theory \cite{Bahramy11}. 
According to the theory, the SOI-induced band splitting near the Fermi level ({\it{ i.e.}} CBM and VBM) is strongly affected by the unusual electronic structures around its band gap; 
most importantly, the band-gap size and the same symmetry characters of CB and HVB due  to the 
negative crystal field splitting of HVB. 
In BiTeI, the VBM is mainly composed of Te $5p$ orbital whereas the CBM is characterized by Bi $6p$.
However, both the sets of bands share the same symmetry character within $C_{3v}$ space group, thereby 
leading to non-vanishing matrix elements, coupling them through the second-order perturbative $k\cdot p$
hamiltonian.
The fairly narrow gap realized in BiTeI ($\sim $0.36 eV from optical study \cite{Lee11})
combined with the large atomic SOI of Bi further enhances this coupling. 
These microscopic electronic structures should be also taken into account for designing the SOI effect in bulk compounds, besides the inversion-asymmetric crystal structure and the atomic spin-orbit coupling.

 Regarding the minimum band gap, it is estimated to be $\sim 260\, \pm \, 50 $ meV at $k_{z}=\pi/c$ from the SX-ARPES data [Fig. 3(a)]. 
Coincidentally it is quite close to the calculated value ($\sim0.28$ eV), but considerably smaller compared to the result of optical spectroscopy showing the minimum band gap of $\sim 0.36$ eV \cite{Lee11}.
This discrepancy between the bulk may be arising from the modification of the electronic structure in the sub-surface ($\lsim$2 nm) region. 
It can be qualitatively confirmed from the depth of the CBM. From the band calculation, $E_{\rm{F}}$ is expected to be around $142$ meV above the CBM, if we assume that the Hall number $n_{\rm{H}}=4.5{\times} 10^{19}$ cm$^{-3}$ correctly represents the bulk electron density. 
The present SX-ARPES data, nevertheless, indicates $E_{\rm{CBM}}\sim 220$ meV, which is deeper than the calculation by $80$ meV. 
This difference may suggest that SX-ARPES data probes the 3D band dispersions from the boundary region between bulk and sub-surface (the probing depth of $h\nu = 550$ eV ARPES is typically $\sim 1$ nm), which is slightly shifted downward by the band bending effect  \cite{Ishizaka11}. 
Recently, the formation of the charge accumulation layer and the resulting band-gap narrowing have been reported in CdO \cite{King10}. Similar situation may be also realized in the sub-surface region of BiTeI. 

Finally we would like to discuss the topology of the 3D Rashba-split bands. 
The calculated Fermi surface ($E_B=0$) and the constant-energy-surface (CES) of HVB at $E_B=0.58$ eV are shown in Fig. 4(b,c). 
$E_B=0.58$ eV corresponds to $100$ meV below the VBM ($E_{\rm{VBM}} \sim 0.48$ eV) and $70$ meV above the band crossing point ($E_{B} \sim 0.65$ eV), as shown in Fig. 4(a). 
Since the ARPES intensity of the CB forming the Fermi surfaces is very weak, here we focus on the CES of HVB at $E_B=0.58$  [Fig. 4(c)]. 
This CES indicates a doughnut-like topology with the strong trigonal deformation, which is represented by a set of two separate distorted ellipsoids of origin-symmetry, along $\Gamma$-A-L-M plane. 
This topology of CES reflects the 3D Rashba-type splitting of VB as a consequence of SOI in $P3m1$ symmetry. 
To confirm this experimentally, we plot in Fig. 4(d) the $k_z$-dependent SX-ARPES intensity at $E_{{B}}=0.58$ eV with the energy window of $\pm 4$ meV. 
The upper region of the image ($\pi/c < k_{z} <2\pi/c$) is obtained by symmetrizing the lower region ($0 < k_{z} < \pi/c$), which corresponds to the data recorded using $h\nu = 540 \sim 585$ eV. 
The distribution of intensity, centered at A-point, is in a good correspondence with the cross-section view of calculated CES (indicated by red curves). 
It thus shows the direct evidence of the band topology with the 3D-Rashba-type splitting, as shown in Fig. 4(c), indeed realized in the bulk BiTeI. 
Since the Fermi energy lies slightly above the band crossing point of the conduction band, the actual Fermi surfaces in the bulk should be consisting of two components, inner and outer ones as shown in Fig. 4(b). In its cross-section view along $\Gamma$-A-L-M plane, two distorted ellipsoids slightly intersect with each other in the vicinity of A point. 
Thus, the magneto-transport properties of this compound should be dominated by the electrons forming such 3D spin-split inner and outer Fermi surfaces. The recent optical study on series of dopant-introduced BiTeI shows that the chemical potential of this compound can be continuously varied from CBM to above the crossing point \cite{Lee11}. 
Together with this feasibility of carrier control, BiTeI possesses the potential to realize various spin-dependent transport/optical properties \cite{Hirsch99} and spintronics devices \cite{Manchon08,Chernysnov09,Miron10}, as well as the new type of SOI-induced unconventional superconductivity \cite{Cappelluti07,Bauer04,Togano04}.

In summary, we investigated the Rashba-type spin splitting in bulk of BiTeI using $h\nu$-dependent SX-ARPES.
Band dispersions indicated a moderate $k_z$-dispersion with the minimum band gap at $k_z = \pi/c$, showing a good agreement with the relativistic first-principles calculations.
The conduction band minimum and the valence band maximum located within the $k_z = \pi/c$ plane both showed a huge splitting, thus offering the direct evidence of giant Rashba-type spin splitting in bulk.
The constant-energy surface at the valence band top indeed showed the doughnut-like topology, confirming the 3D Rashba-split Fermi surface realized in the bulk BiTeI. 
These results show the potential of BiTeI to realize the strong spintronic functions.

This work was partly supported by JST, PRESTO, and by JSPS through the ``Funding Program for World-Leading
Innovative R\&D on Science and Technology (FIRST Program)".
The synchrotron radiation experiments were performed at BL17SU in SPring-8 with the approval of RIKEN (Proposal No. 20100055).

\newpage

\begin{figure}[htbp!]
\caption{
(color online) (a) Crystal structure and (b) Brillouin zone of BiTeI.
(c) Momentum cuts of measurement in $k_{x}-k_{z}$ ($\Gamma$-A-L-M) plane, corresponding to the photon energies of $h{\nu}=530\sim 595$ eV (5 eV steps). Those recorded by energy-resolution of ${\Delta}E=100$,(150) meV are shown by solid (broken) curves.
\label{fig1}
}
\end{figure}

\begin{figure}[htbp!]
\caption{
(color online)
(a)-(c) SX-ARPES images along $k_x$ direction by using $h{\nu}=540, 565,$ and  $585$ eV,  respectively. 
Black circle markers with error bars indicate the peak positions of the SX-ARPES intensity. 
The green rectangle in (c) shows the CBM and VBM located around  A point. (d) Out-of-plane band dispersions along $\Gamma$-A line obtained by SX-ARPES  using $h{\nu}=530$ ${\sim}$ $590$ eV, which correspond to $k_{z}=-0.22{\pi/c}$ ${\sim}$ $1.11{\pi/c}$. (e)-(g) Band dispersions along $k_x$ direction obtained by calculation corresponding to $k_{z}=0$, $0.56{\pi/c}$, and ${\pi/c}$, respectively. 
(h) Band dispersions along $\Gamma$-A line obtained by calculation.
\label{fig2}}
\end{figure}

\begin{figure}[htbp!]
\caption{
(color online) 
 (a) SX-ARPES image around the CBM and the VBM at $k_{z}=\pi/c$ ($h{\nu}=585$ eV). Open circle (square) markers on the HVB are the peak positions of the EDCs (MDCs), whereas the broken curves overlaid on the CB are guides for eyes.
The corresponding EDCs are shown in (b).
(c) The band calculation near the Fermi level at $k_z=\pi/c$ along $k_x$ direction, showing the CBM and VBM. 
\label{fig3}
}
\end{figure}

\begin{figure}[htbp!]
\caption{
(color online)
Calculated band dispersion at $k_z=\pi/c$ (a), 3D views of Fermi surface ($E_B=0$) (b), and constant energy surface (CES) at $E_{{B}}=0.58$ eV (c).
(d)  SX-ARPES intensity image at $E_{{B}}=0.58$ eV in $k_{x}-k_{z}$ plane, which are recorded by using $h{\nu}=540$ $\sim$ $585$ eV. 
The right inset (the red curves in the main panel) indicates the calculated (cross-section) view of CES at $E_B=0.58$ eV.  
\label{fig4}
}
\end{figure}

\newpage
Fig. 1 M. Sakano {\it{et al}}.
\begin{center}
 \includegraphics[height=12cm,width=12cm,keepaspectratio,clip]{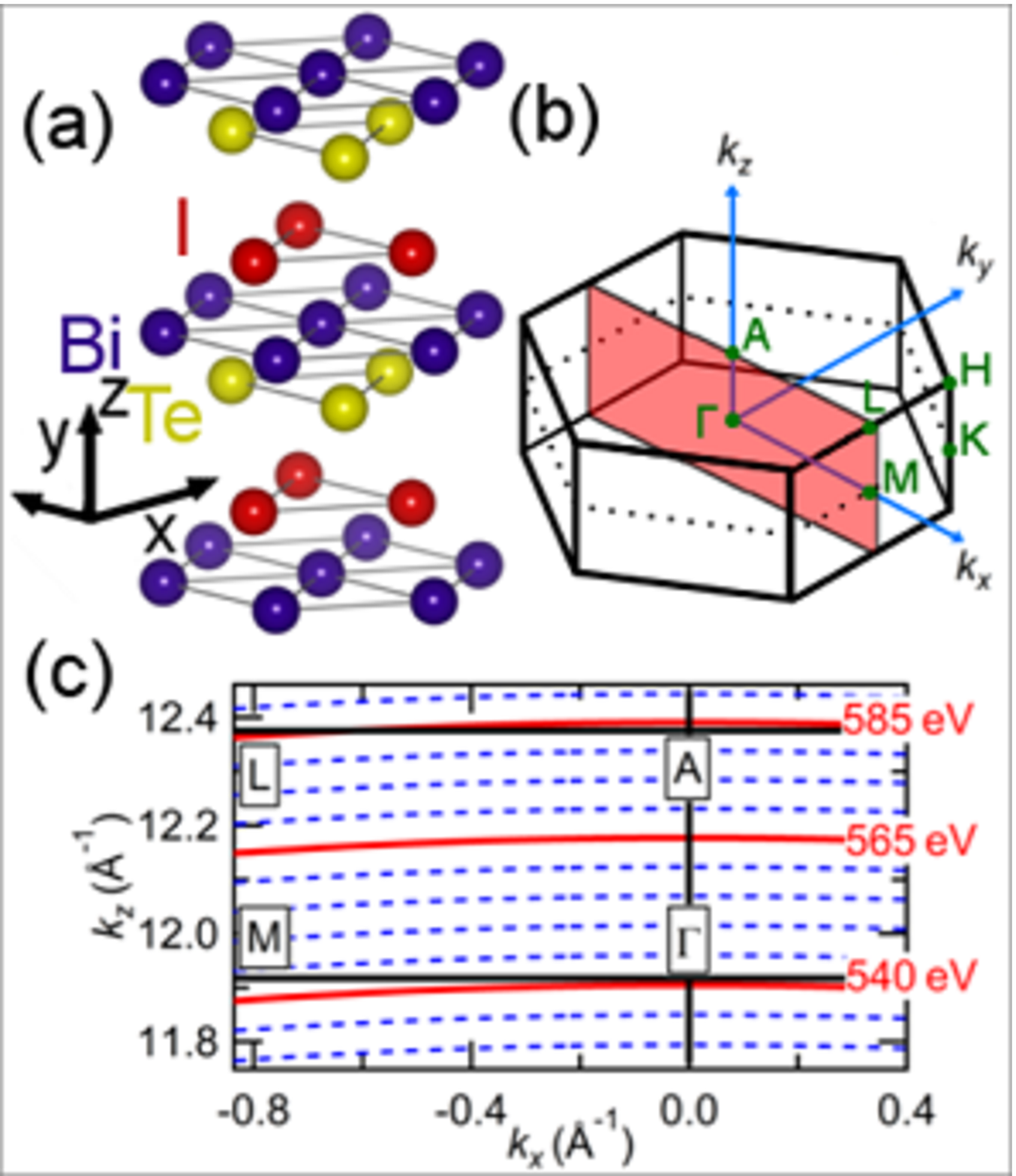}
\end{center} 

\newpage
Fig. 2 M. Sakano {\it{et al}}.
\begin{center}
 \includegraphics[height=14cm,width=14cm,keepaspectratio,clip]{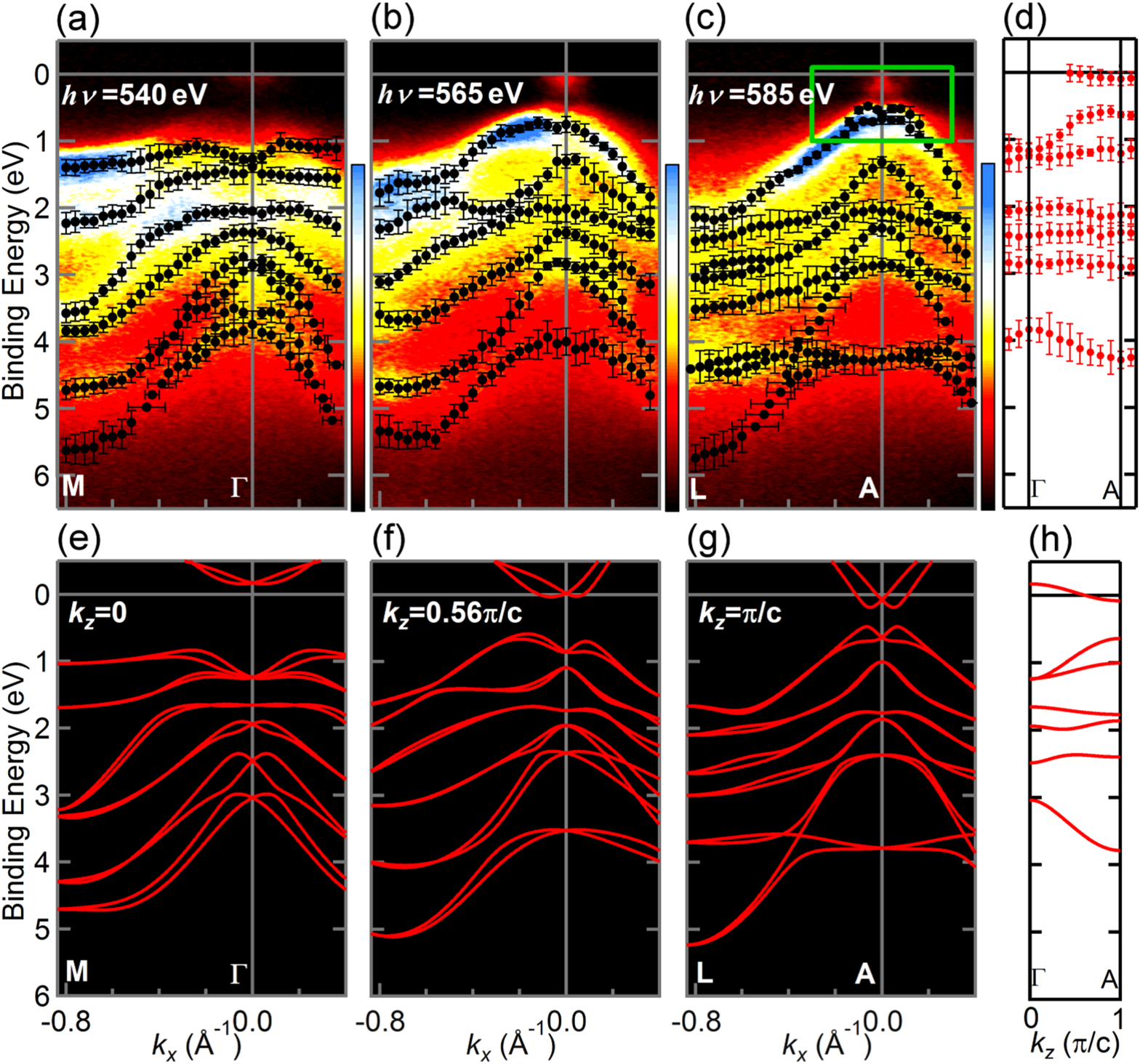}
\end{center} 

\newpage
Fig. 3 M. Sakano {\it{et al}}.
\begin{center}
 \includegraphics[height=12cm,width=12cm,keepaspectratio,clip]{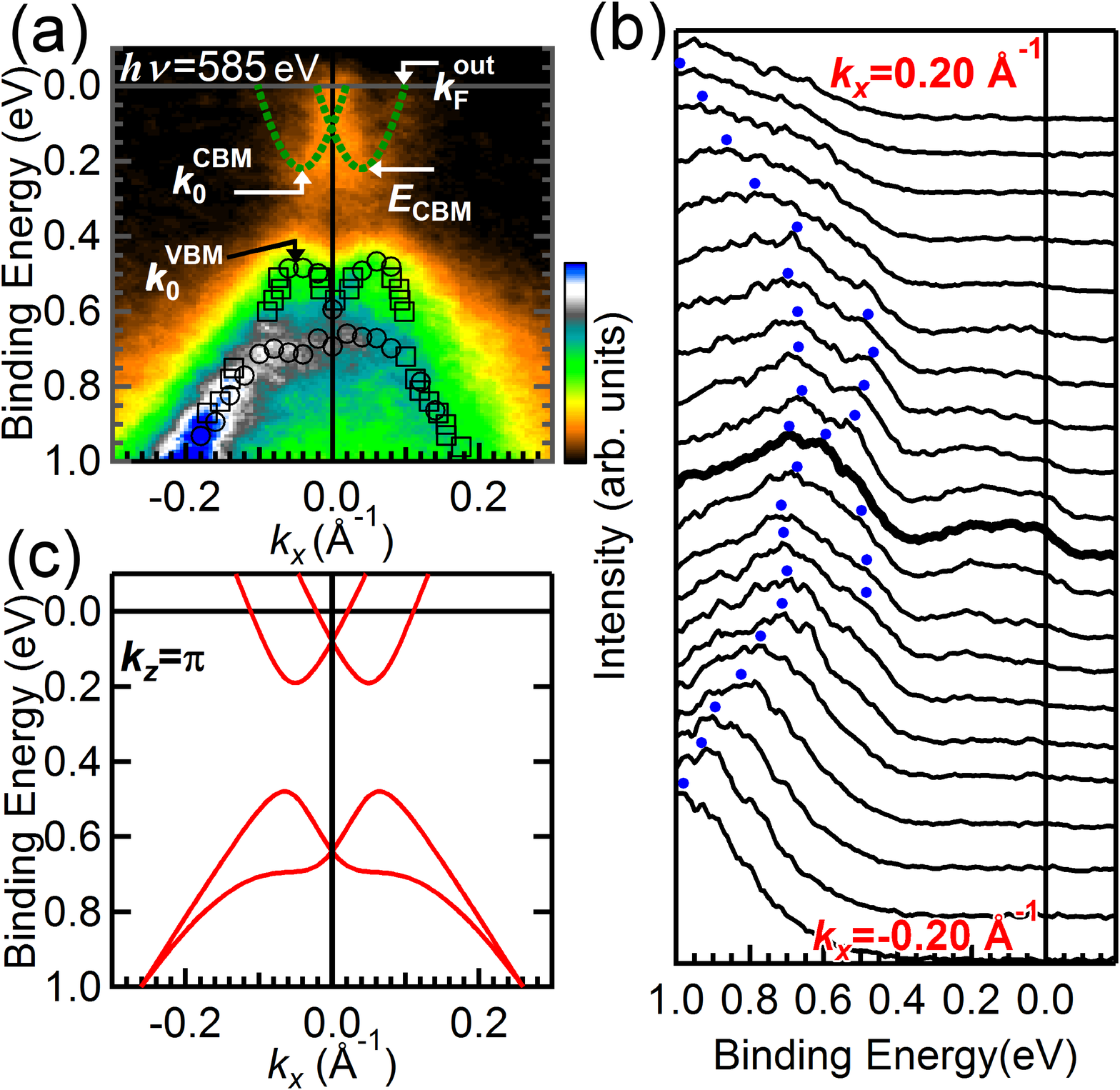}
\end{center} 

\newpage
Fig. 4 M. Sakano {\it{et al}}.
\begin{center}
 \includegraphics[height=12cm,width=12cm,keepaspectratio,clip]{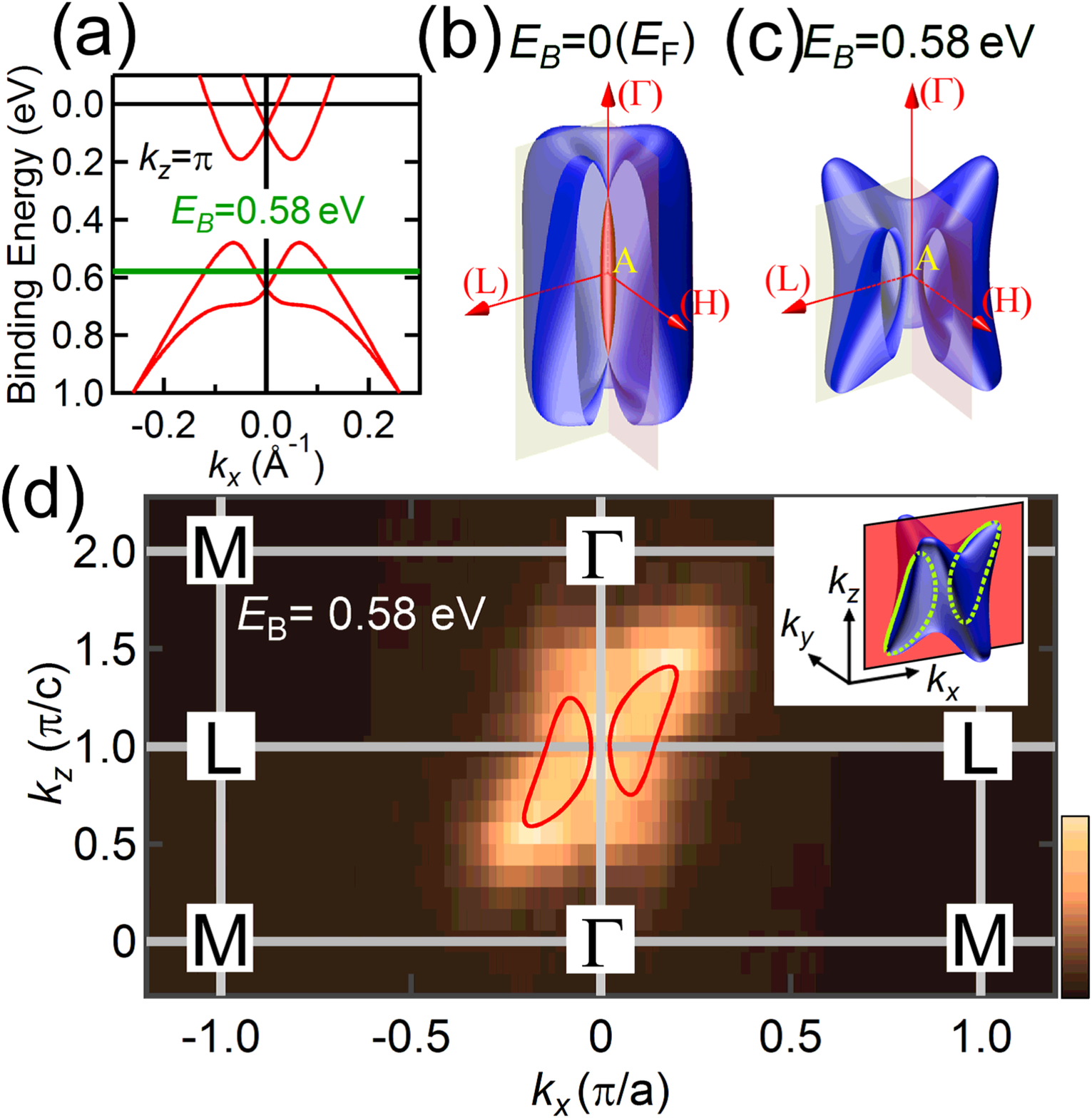}
\end{center}


\begin{thebibliography}{99}
\bibitem{comment1}
The details of first-principles calculations can be found in \cite{Bahramy11}.
\end{thebibliography}
\end{document}